# Work Function Engineered Charge Plasma-Germanium Double Gate Tunnel Field Effect Transistor for Low-Power Switching Applications

Sambhu P. Malik, Ajeet K. Yadav, Robin Khosla*

Department of Electronics and Communications Engineering, National Institute of Technology, Silchar, 788010, Assam, India

**Abstract**
Here, we propose a Charge Plasma (CP)-based Germanium Double Gate Tunnel Field-Effect Transistor (Ge-DGTFET) device structure, where a CP is induced in the heavily doped source region using the work function engineering of source electrode. The CP enables creation of electrical metallurgical junction and converts n-p-n to p-n-p-n structure of TFET and enhances the drain current, reliability, eliminate additional pocket ion-implantation. The proposed CP-Ge-DGTFET device structure revealed excellent electrical DC performance as compared to the conventional Ge-DGTFET device structure such as high ON current ($I_{ON}$), excellent $I_{ON}/I_{OFF}$ ratio, and low sub-threshold swing of ~4.7×10$^{-4}$ A/μm, ~1.8×10$^9$, and ~5.23 mV/dec, respectively. Furthermore, analog/RF analyses revealed high transconductance, upright cut-off frequency, low overall capacitance, transit time, and power delay product. Therefore, the proposed CP-Ge-DGTFET device structure with alternate channel material Ge, High-κ $Al_2O_3$, and work function engineered CP in source region furnishes high performance and cost-effective solution for next-generation energy-efficient switching applications.

## 1. INTRODUCTION

The advancement of Integrated Circuits (IC) technology has been majorly viable by the incessant downscaling of device dimensions to achieve high performance electronic devices with fast switching speed, higher integration density, low power consumption at reduced cost per device. However, the scaling driven performance improvement of silicon-based devices is expected to end soon as the devices dimensions reach physical limits and leads to short channel effects [1]. In addition, supply voltage scaling is the foremost challenge that leads to devices high power consumption due to fundamental thermionic limit defined by Boltzmann's Tyranny (i.e. minimum sub-threshold swing ($S$) of 60 mV/dec is needed to switch the transistor) [2], [3]. In this regard, investigation of new device structures and integration of alternate materials in devices are expected to provide a solution to performance improvement of next-generation complementary-metal-oxide-semiconductor (CMOS) technology [4-8].

The Tunnel Field-Effect Transistor (TFET) is a vital contender for CMOS technology thanks to the capability to condense sub-threshold swing ($S$<60 mV/dec) and supply voltage ($V_{DD}$<1V) enabled by quantum mechanical band-to-band tunneling (BTBT) principle [9]. The p-n-p-n device structure of TFET with pocket region amid the source & channel delivers high performance and energy efficient switching in comparison to TFET's conventional p-i-n structure due to reduced tunnel width and boosted lateral electric field [10],[11]. However, additional ion-implantation process other than source & drain is required to create the pocket region during device fabrication process that raises the complexity, reduced gate control, & manufacturing cost [12].

In this work, three major modifications are proposed in conventional p-i-n TFET, (i) Integration of Ge as an active semiconductor material due to its low bandgap, high electron & hole mobility, minimize heterostructure interface defects and to maintain the ease of device processing [13]; (ii) High work function electrodes in the source region to electrically induce hole plasma in the metallurgical junction, to eliminate the surplus pocket ion-implantation [14-16]; (iii) High-κ $Al_2O_3$ as a gate dielectric due to its excellent insulator properties and better compatibility with Ge [17-18]. The electrical DC & Analog/RF characteristics are considered.

## 2. DEVICE STRUCTURE AND METHODS

The cross-section view of conventional and proposed TFET device structures investigated in this work are presented in Fig. 1(a) and 1(b), respectively. In conventional and proposed TFET device structures, the source, channel, and drain regions are chosen to be Ge semiconductor material to minimize the heterostructure defects that can be introduced due to lattice mismatch between different semiconductors. Here, High-κ $Al_2O_3$ is used as a gate dielectric due to its excellent material properties, such as better compatibility with Ge, wide bandgap (~8.7 eV), moderate dielectric constant (~9), good temperature & kinetic stability, high-quality atomic layer deposition processing and lower number of defects [17-18]. For electrical gate contacts, Ti with work function ($\phi_m$ of ~4.33 eV) is used due to CMOS compatibility, low cost and ease of processing. Moreover, 22 nm physical gate length ($l_g$), 100 nm source length ($l_s$) & drain length ($l_d$), and 2 nm gate dielectric thickness ($t_{ox}$) to maintain ~1 nm effective oxide thickness (EOT) are used in both the investigated device structures. In the conventional Ge-DGTFET device structure, p$^{++}$(1×10$^{20}$ cm$^{-3}$), p$^{-}$(1×10$^{16}$ cm$^{-3}$), and n$^{+}$(1×10$^{18}$ cm$^{-3}$) doping is introduced in the source, channel and drain regions, respectively, to form TFET's conventional p-i-n structure for comparison and calibration [19]. Though, in the proposed Charge Plasma (CP) based Ge-DGTFET device structure, n$^{++}$ (1×10$^{20}$cm$^{-3}$), p$^{-}$ (1×10$^{16}$cm$^{-3}$), and n$^{+}$ (1×10$^{18}$cm$^{-3}$), doping is introduced in the source, channel and drain regions, respectively. Also, the source is heavily doped as compared to the drain to extend the drain depletion region that mitigates off-state current and minimizes the ambipolarity [20]. Further, CP can be induced electrically in source region provided: (i) work function of electrode ($\phi_m$) is different as compared to the sum of semiconductor electron affinity ($\chi_s$) and half of energy band gap ($E_g$), related using (1):

$$\phi_m \neq \chi_S + \left(\frac{E_g}{2}\right) \quad (1)$$

and (ii) thickness of semiconductor ($t_s$) is equal to or less than the Debye length ($l_d$), which becomes crucial for nanoscale devices [14], related using (2):

$$t_S \leq l_d \quad (2)$$

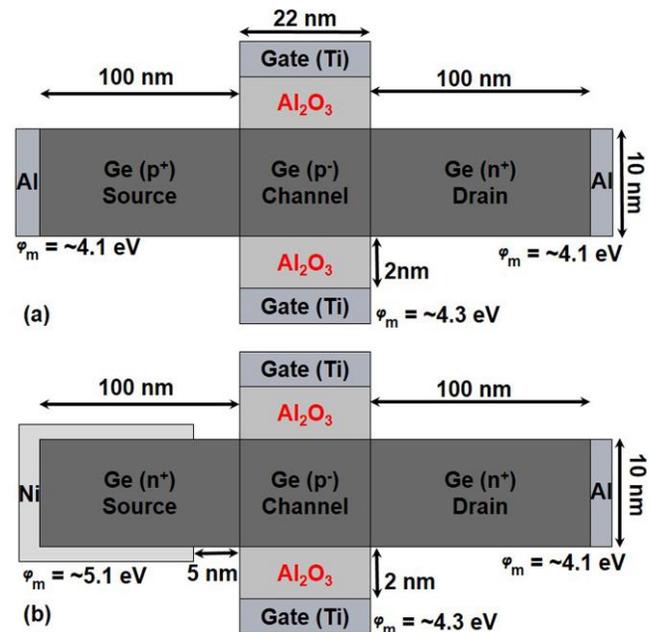

**Fig. 1.** Crossection view of investigated TFET device structures: (a) Conventional DG-TFET (b) Proposed Charge Plasma based Ge-DGTFET.

Recently, integration of Ni metal contact over Ge has proven formation of NiGe on thermal treatment and oxidation resistance to form good metal-semiconductor schottky contact [6, 21]. Thus, high work function ($\phi_m$ of 5.1 eV) source electrode (Ni) is used to create excess hole plasma in the n-region electrically so as to form p-n-p-n structure of the CP-Ge-DGTFET device structure. Additionally, a noteworthy gap of ~5 nm is maintained amid the source (plasma) and gate electrodes to create an electrical metallurgical junction and n-region amid channel and hole plasma.

The simulation results of investigated device structures are performed using ATLAS device simulator of Silvaco [22]. In this work, the non-local BTBT model is most crucial and used for TFET simulations due to the BTBT based principle mechanism of TFET. The Shockley-Read-Hall (SRH) model estimates the generation and recombination of charge carriers. The bandgap narrowing (BGN) and Fermi-Dirac Distribution (FDD) models are used for heavily doped regions and carrier statistics, respectively. The drift-diffusion and Poisson equations are solved after discrete meshing of the device structure using the drift-diffusion current model that relates to the transport of the charge carriers. The Lombardi mobility model is utilized for the field-dependent mobility and concentration effects [22], [23]. The software has been calibrated with data reported in [19], as shown in Fig. 2.

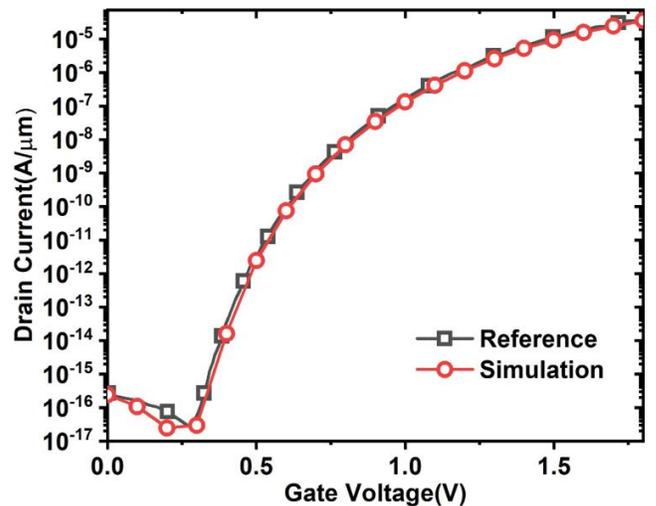

**Fig. 2.** Model calibration with the reference work reported in [19].

## 3. RESULTS AND DISCUSSIONS

Fig. 3 shows the creation of hole plasma which divulges the carriers density along channel surface of the proposed CP-Ge DGTFET device structure. The corresponding contour plot is displayed in Fig. 3(a), which clearly reveals that the source region's excess hole carrier concentration (hole plasma) has been developed and thus results in electrical creation of TFET's p-n-p-n structure. Further, for quantitative analysis of hole plasma, a "cut-line" is drawn 2 nm below the $Al_2O_3$/Ge interface of the contour plot to obtain the electron-hole carrier density vs. lateral distance plot, as presented in Fig. 3(b). An excess hole carrier concentration of ~$1.40 \times 10^{20}$ cm$^{-3}$ in the source region just below the plasma electrode (labeled with green-color spheroid in Fig. 3(b)) confirms that the hole plasma has been created within the source region. Also, since a small gap ($L_{gap}$ of ~5 nm) is kept between the plasma electrode and gate electrode, thus, a small n-type region still remains between some portion of source (p-type due to hole plasma) and channel (p$^-$) regions. As a result, metallurgical junction is created electrically and an abrupt energy band is developed close to source-channel interface in source. Thus, n-p-n device structure transforms to TFET's p-n-p-n structure.

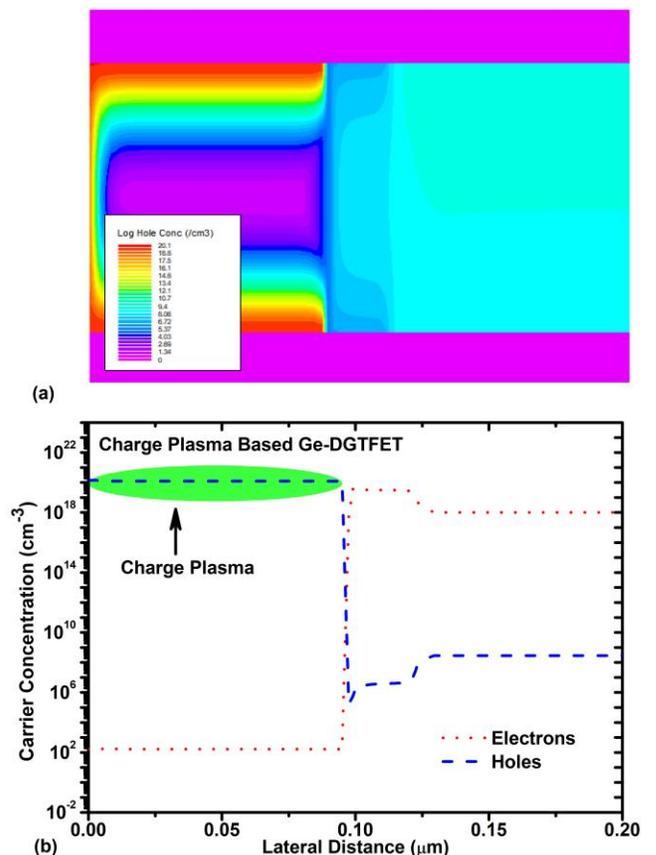

**Fig. 3.** (a) Creation of hole plasma concentration extracted using contour plot. (b) Electron-hole carrier density along the "cut-line" (drawn 2nm below the high-κ $Al_2O_3$/Ge interface).

### 3.1 DC Analysis

Fig 4(a) shows the $I_D$ vs. $V_{GS}$ (Transfer) characteristics of Ge-DGTFET and the proposed CP-Ge-DGTFET device structure. The ON current extracted from the transfer characteristics of Ge-DGTFET and CP-Ge-DGTFET is ~$1.02 \times 10^{-4}$ A/μm and ~$4.7 \times 10^{-4}$ A/μm at $V_{GS}$ of 1V, respectively. Further, the OFF current is revealed to be ~$3.27 \times 10^{-12}$ A/μm and ~$8.5 \times 10^{-13}$ A/μm at gate voltage ($V_{GS}$) of 0V for conventional Ge-DGTFET and CP-Ge-DGTFET, respectively. Thus, CP-Ge-DGTFET offers one order improved OFF-current in comparison to p-i-n Ge-DGTFET. Furthermore, the point sub-threshold swing ($S_{pt}$) is extracted using equation (3) to be ~37.7 mV/dec and ~5.23 mV/dec whereas the average sub-threshold swing ($S_{avg}$) is estimated using equation (4) to be ~51.3 mV/dec and ~29.2 mV/dec for conventional Ge-DGTFET and CP-Ge-DGTFET, respectively.

$$\left(S_{pt}\right)_{V_{GS}} = \left(\frac{dV_{GS}}{d\log(I_D)}\right)_{V_{GS}} \quad (3)$$

$$S_{avg} = \frac{V_T - V_{OFF}}{\log(I_{VT}) - \log(I_{OFF})} \quad (4)$$

Therefore, the proposed CP-Ge-DGTFET shows better electrical characteristics as compared to conventional Ge-DGTFET device structure and investigated for further electrical characteristics.

Fig. 4(b) shows the Output ($I_D$-$V_{DS}$) characteristics of the proposed CP-Ge-DGTFET device structure with variation in $V_{DS}$ from 0V to 1V at various fixed $V_{GS}$ of 0.5V, 0.7V and 1V. The drain current is observed to escalate from ~$9.8 \times 10^{-5} A/\mu m$ to ~$1.7 \times 10^{-3} A/\mu m$ with variation in $V_{GS}$ from 0.5V to 1V, at fixed $V_{DS}$ of 1V. Additionally, at fixed $V_{GS}$ of 1V, when $V_{DS}$ is 0V to ~0.4V, the drain-channel junction barrier is reasonably high that consequences to low charge carrier's transport from channel to the drain and thus subordinate drain current called the "tunnel resistance dominated region". Further, with variation in $V_{DS}$ from ~0.4V to ~0.8V, the drain-channel junction barrier decreases, that accounts to exponential increase of drain current named the "Channel resistance dominated region". However, with further increase in $V_{DS}$ from ~0.8V to ~1.0V, the drain current tends to saturate due to reduced concentration of carriers in the channel termed the "Saturation Region" [24].

To support the tunnelling mechanism in the proposed CP-Ge-DGTFET device structure the energy band profile needs to be investigated. Fig. 5 demonstrates the energy band profile for CP-Ge-DGTFET device structure in the ON and OFF states. Here, the "OFF" state is defined at $V_{GS}$ of 0V and $V_{DS} \leq 0.4V$, where source's valance band is significantly separated from channel's conduction band owing to wide tunnel barrier and results in OFF state current. Alternatively, "ON" state is defined at $V_{GS}$ of 1V and $V_{DS}$ of 0.4V, where channel's conduction band is aligned with source's valence band, thus minimises tunnel width, and probability of carriers tunnelling increases. The TFET's ON-current depends on tunnelling probability T(E) & related using equation (5) [25]:

$$T(E) = \exp\left(\frac{-4\lambda\sqrt{2m^*}E_g^{3/2}}{3q\hbar(E_g + \Delta\phi)}\right) \quad (5)$$

where $\lambda$, $E_g$, $m^*$, $\Delta\Phi$, q, and $\hbar$ is the screening tunnelling length, bandgap of semiconductor, effective carrier mass, energy range where tunnelling occurs, charge of electron, and reduced Planck's constant, respectively. Equation (5), marks that small bandgap, and low effective mass material enhances the BTBT rate (tunnelling probability). Thus, as compared to Si, the Ge-based DGTFET is expected to show a high $I_{ON}$. In addition, tunnelling screening length ($\lambda$) has a noticeable impact on TFET performance, related using equation (6) [25]:

$$\lambda = \sqrt{\frac{\epsilon_s}{\epsilon_{ox}} t_{ox} t_s} \quad (6)$$

where $\epsilon_s$, $\epsilon_{ox}$, $t_{ox}$ and $t_s$ is the relative permittivity of semiconductor channel material, relative permittivity of gate dielectric, thickness of the gate dielectric, and thickness of the channel material, respectively. Thus, tunnelling probability can be augmented on reduction of $\lambda$ by using an alternate thin high-κ gate oxide ($Al_2O_3$) due to formation of strongly modulated channel bands.

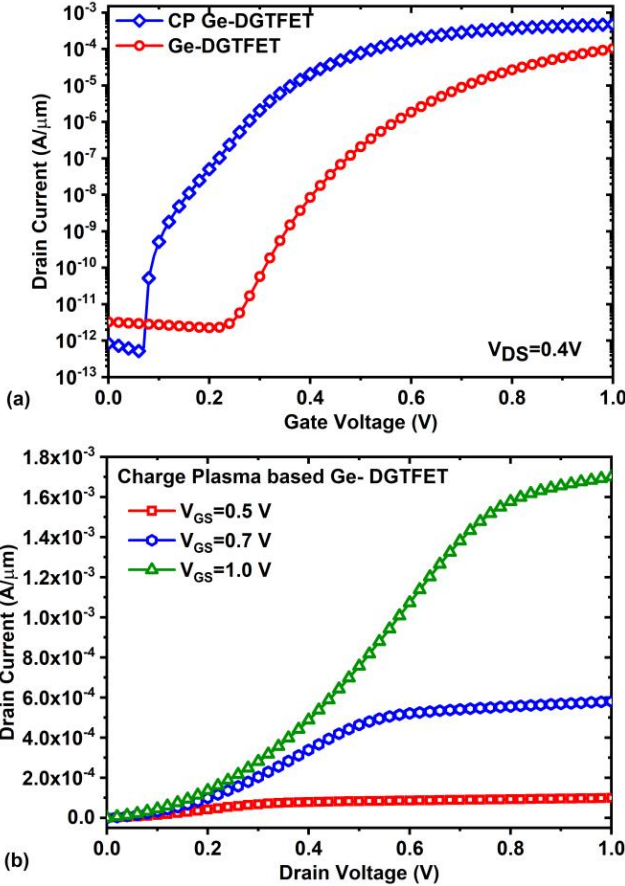

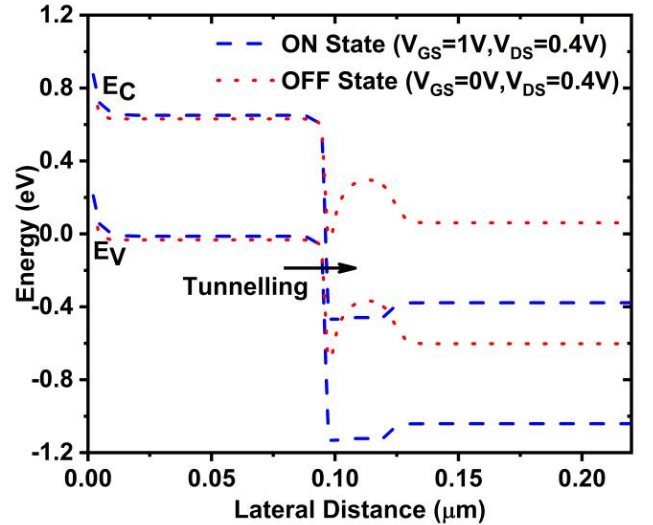

**Fig. 4.** (a) Transfer Characteristics ($I_D$-$V_{GS}$) of the investigated conventional Ge-DGTFET and proposed CP-Ge-DGTFET device structure. (b) The output characteristics ($I_D$-$V_{DS}$) of the proposed CP-Ge-DGTFET device structure with variation in Gate Voltage ($V_{GS}$).

**Fig. 5.** Energy band profile for CP-Ge-DGTFET device structure in ON and OFF states.

Further, to demonstrate how source work function engineering can improve the device performance, Fig. 6 depicts the electric fields of CP-Ge-DGTFET and Ge-DGTFET in the ON state ($V_{GS}$ of 1V and $V_{DS}$ of 0.4V). It is observed that there is a sharp upsurge in the electric field at the source-channel junction and a reduction at the channel-drain junction, attributes to dominant tunneling at source-channel junction results in high $I_{ON}$ in agreement with equation (5).

### 3.2 Analog/ RF Analysis

The Analog/RF characteristics of CP-Ge-DGTFET device structure is evaluated, where a 100kHz ac throughput signal is applied to the device structure and various key parameters like parasitic capacitances ($C_{gs}$, $C_{gd}$, $C_{gg}$), transconductance ($g_m$), cut-off frequency ($f_T$), transit time ($\tau$), efficiency, and power delay product (PDP) are analysed.

The parasitic capacitances ($C_{gs}$, $C_{gd}$, $C_{gg}$) related with the devices play a critical role in analysing its analog behaviour while operating at high frequencies. In fact, the stored charge density in gate, drain, and source are obtained from parasitic capacitances. Over and above, they form a path between input and output that creates circuit oscillations, signal delay, and power dissipation [26]. Fig. 7 presents the Capacitance ($C_{gs}$, $C_{gd}$, $C_{gg}$) vs. gate voltage characteristics of the CP-Ge-DGTFET device structure with variation in $V_{DS}$ of 0V (Fig. 7(a)) and 0.4V (Fig. 7(b)). Here, $C_{gs}$ is estimated to be ~0.004 fF and ~1.04 fF that is much smaller than the assessed $C_{gd}$ value of ~7 fF and ~6.49 fF at $V_{DS}$ of 0V and 0.4V, respectively. Therefore, gate capacitance ($C_{gg}$) of CP-Ge-DGTFET is analogous to $C_{gd}$ evident from Fig. 7. The overall $C_{gg}$ is related as:

$$C_{gg} = C_{gs} + C_{gd} \quad (7)$$

Moreover, $C_{gg}$ increase rapidly with increase of $V_{GS}$ from 0V to 1V due to electrons aggregation in the channel that results in increase of capacitance [26]. Furthermore, a right shift in the $C_{gg}$ curve is observed with variation in $V_{DS}$ from 0V (Fig. 7(a)) to 0.4 V (Fig. 7(b)), probably due to higher

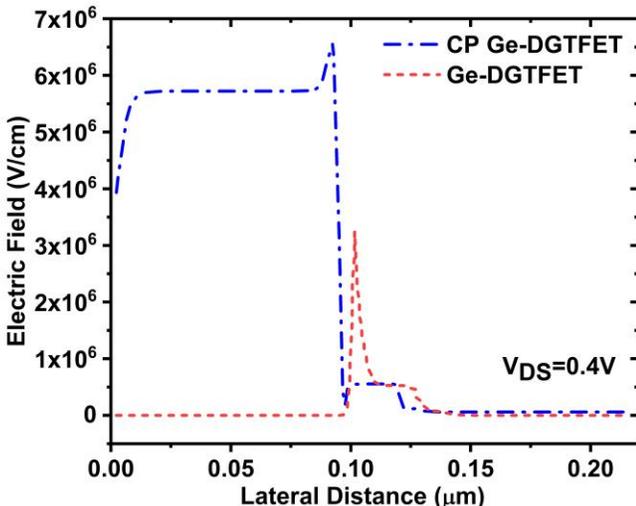

**Fig. 6.** Electric field profile of CP-Ge-DGTFET and conventional Ge-DGTFET device structures.

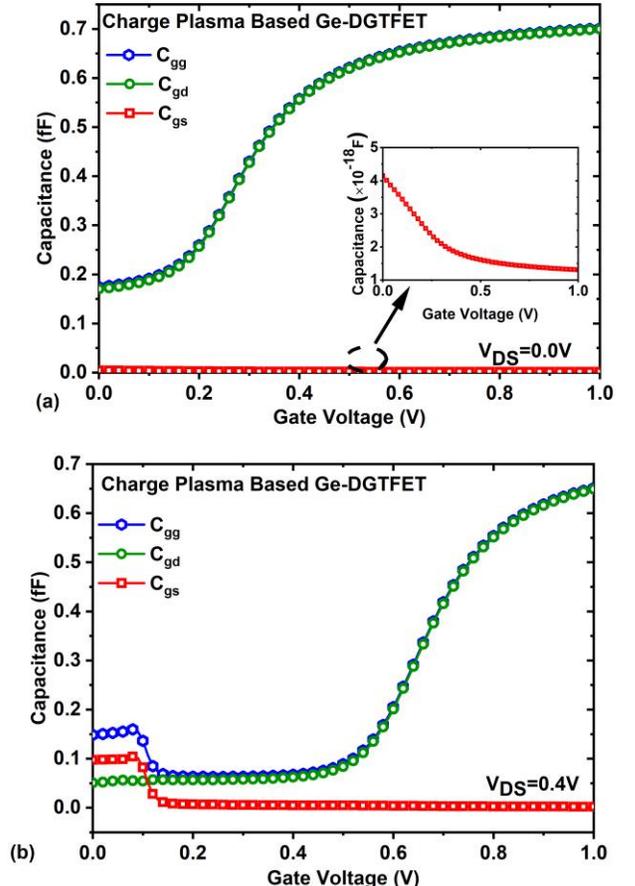

**Fig. 7.** Capacitance ($C_{gs}$, $C_{gd}$, and $C_{gg}$) vs. Gate Voltage Characteristics of CP-Ge-DGTFET at $V_{DS}$ of (**a**) 0V and (**b**) 0.4V.

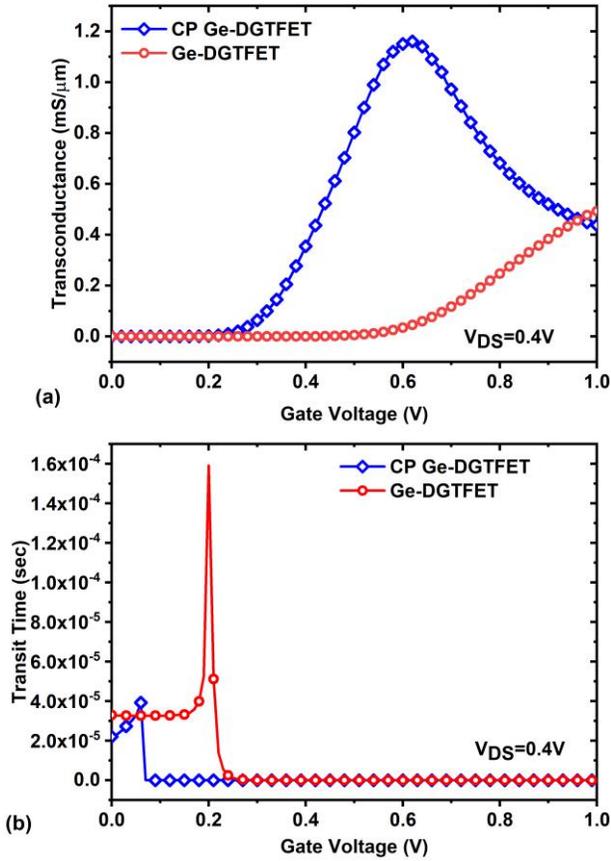

$$GBW = \frac{g_m}{2\pi 10 C_{gd}} \quad (10)$$

The estimated maximum $f_T$ and GBW of the CP-Ge-DGTFET device structure is ~1.44 THz and ~144 GHz, respectively which is much higher than ~129 GHz and ~12.9 GHz, respectively for conventional Ge-DGTFET, reveals CP-Ge-DGTFET exceptional candidate for RF applications.

Moreover, the transit time ($\tau$) is also a chief metric for RF analysis which defines the interval of charge carrier's transition from source to drain. A high-speed device has a low transit time, and is inversely proportional to $f_T$, as follows:

$$\tau = \frac{1}{2\pi f_T} \quad (11)$$

Fig. 8(b) depicts the transit time characteristics with variation in gate voltage from 0V to 1V at fixed $V_{DS}$ of 0.4V for conventional Ge-DGTFET and CP-Ge-DGTFET device structures. The CP-Ge-DGTFET device structure revealed reasonably low $\tau$ of ~3.9×10⁻⁵ sec as compared to conventional Ge-DGTFET device structure $\tau$ of ~1.59×10⁻⁴ sec. Hence, the proposed CP-Ge-DGTFET device structure must take less time to perform specific operations.

Over and above, the device efficiency is momentous parameter for analog applications defined using horizontal drain electric field that opposes the vertical gate electric field in the channel. Hence, results in effective channel voltage of $V_{GS}$-$V_{th}$-$V_{DS}$. Thus, at higher $V_{DS}$ of 0.4V an effective gate voltage of ~0.6V is required to enable the increase of $C_{gg}$ with $V_{GS}$.

**Fig. 8.** (a) Transconductance vs. $V_{gs}$ Characteristics and (b) Transit time vs. $V_{gs}$ characteristics, with variation in $V_{GS}$ from 0V to 1V at fixed $V_{DS}$=0.4V for conventional Ge-DGTFET & CP-Ge-DGTFET device structures.

Further, the transconductance ($g_m$) is an essential parameter for analog analysis, which exemplifies device's amplification capability and related using equation (8):

$$g_m = \frac{\partial I_D}{\partial V_{GS}} \quad (8)$$

Fig. 8(a) shows the transconductance characteristics with variation in gate voltage from 0V to 1V at fixed $V_{DS}$ of 0.4V for conventional Ge-DGTFET and CP-Ge-DGTFET device structures. The CP-Ge-DGTFET device structure revealed a high peak $g_m$ of ~1.16 mS/µm at $V_{GS}$ of ~0.6V which is much higher and in the operating range (<1V) in comparison to the Ge-DGTFET device structure.

Furthermore, cut-off frequency ($f_T$) and gain bandwidth product (GBW) play a vibrant role for frequency analysis and must be high for RF applications. $f_T$ is related using (9) & for a fixed DC gain (~10), GBW is related using (10) [27, 28]:

$$f_T = \frac{g_m}{2\pi C_{gs}\sqrt{1+2C_{gd}/C_{gs}}} \approx \frac{g_m}{2\pi(C_{gs}+C_{gd})} \approx \frac{g_m}{2\pi C_{gg}} \quad (9)$$

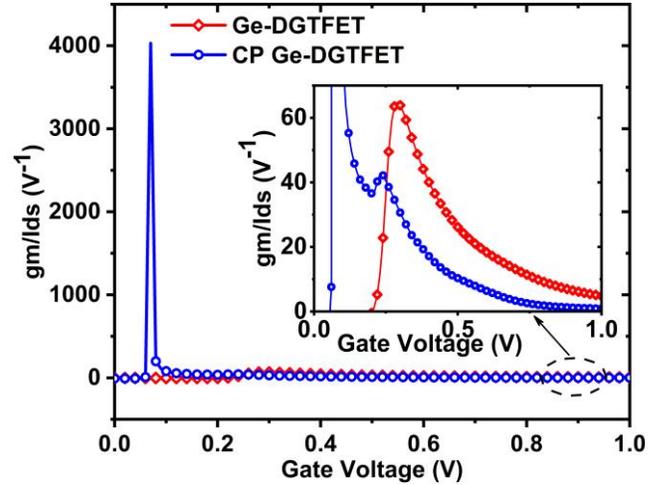

**Fig. 9.** The device efficiency characteristics for conventional Ge-DGTFET and CP-Ge-DGTFET device structures.

transconductance to current ratio ($g_m/I_{ds}$). Fig. 9 shows the device efficiency characteristics for conventional Ge-DGTFET and CP-Ge-DGTFET device structures. The comparative plot of device efficiency clearly depicts greater peak of ~4000 V⁻¹ for CP-Ge-DGTFET device structure, enables it more suitable for the analog/ RF applications.

Furthermore, the requisite switching energy for transistors ON-OFF transition is defined by power-delay product (PDP). The PDP is an important metric for performance evaluation of a low power switching device and related using (12) [29]:

$$PDP = C_{gg} \times V_{DD}^2 \quad (12)$$

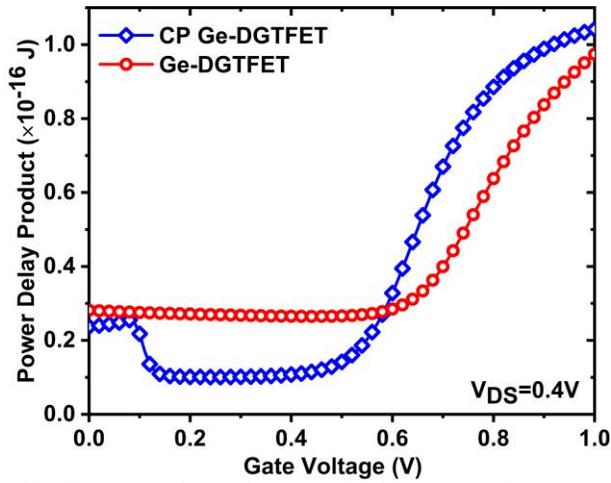

**Fig. 10.** The power delay product vs. gate voltage characteristics at $V_{DS}$ of 0.4V, for conventional Ge-DGTFET and proposed Charge Plasma based Ge-DGTFET device structures.

Fig. 10 shows the PDP-$V_{GS}$ characteristics with variation in $V_{GS}$ from 0V to 1V and at fixed $V_{DS}$ of 0.4V for conventional Ge-DGTFET and CP-Ge-DGTFET device structures. The PDP-$V_{GS}$ characteristics revealed a comparable PDP of $\sim 1.04 \times 10^{-16}$ J and $\sim 9.74 \times 10^{-17}$ J at $V_{GS}$ of 1V for CP-Ge-DGTFET and Ge-DGTFET, respectively. While PDP of CP-Ge-DGTFET device structures is much lower than PDP of conventional Ge-DGTFET device structures for $V_{GS}<0.6V$. Thus, CP-Ge-DGTFET device structure marks low energy dissipation per switching operation remarkable for low-power switching.

**Table 1** Performance comparison of various TFET device structures

| Device Structure | $I_{ON}$ (A/μm) | $V_{GS}$ (V) | S (mV/dec) | $g_m$ (mS/μm) | $f_t$ (Hz) | REF |
|---|---|---|---|---|---|---|
| LTFET | $\sim 1.1 \times 10^{-3}$ | 1.5 | $\sim 11.4$ | $\sim 0.012$ | $\sim 1.1 \times 10^{12}$ | [30] |
| UTFET | $\sim 1.4 \times 10^{-6}$ | 1.0 | $\sim 9.4$ | $\sim 0.006$ | $\sim 3.0 \times 10^{8}$ | [31] |
| DMCG-CPTFET | $8.8 \times 10^{-5}$ | 1.5 | - | - | $\sim 27 \times 10^{9}$ | [32] |
| SiGe-CPTFET | $8.8 \times 10^{-6}$ | 1.0 | $\sim 20$ | $\sim 0.06$ | - | [11] |
| Ge-DGTFET | $\sim 1.0 \times 10^{-4}$ | 1.0 | $\sim 37.7$ | $\sim 0.49$ | $\sim 1.3 \times 10^{11}$ | This Work |
| CP-Ge-DGTFET | $\sim 4.7 \times 10^{-4}$ | 1.0 | $\sim 5.23$ | $\sim 1.16$ | $\sim 1.4 \times 10^{12}$ | This Work |

Finally, Table 1 summarizes performance of various devices structures as compared to this work & clearly reveals high performance of proposed CP-Ge-DGTFET device structure.

## 4. CONCLUSIONS

In Summary, Charge Plasma (CP)-Ge-DGTFET device structure is designed and simulated, where a CP is induced in heavily doped source region using work function engineered source, high-mobility Ge is used as alternate semiconductor and $Al_2O_3$ as high-κ gate dielectric. The CP-Ge-DGTFET device structure revealed excellent electrical dc performance compared to conventional Ge-DGTFET device structure such as high ON current ($I_{ON}$), excellent $I_{ON}/I_{OFF}$ ratio, and low sub-threshold swing of $\sim 4.7 \times 10^{-4}$ A/μm, $\sim 1.8 \times 10^{9}$, and $\sim 5.23$ mV/dec, respectively. Furthermore, analog/RF analyses revealed high transconductance, upright cut-off frequency, low transit time and power delay product of $\sim 1.16$ mS/μm, 1.44 THz, $\sim 3.9 \times 10^{-5}$ sec, and $\sim 1.04 \times 10^{-16}$ J, respectively. Therefore, the proposed CP-Ge-DGTFET device structure furnishes high performance and cost-effective solution for next-generation energy-efficient switching applications.

**Declaration of competing interest**

The authors declare that they have no known competing financial interests or personal relationships that could have appeared to influence the work reported in this paper.

**Acknowledgment**

The authors thank National Institute of Technology Silchar, for providing access to Silvaco Atlas TCAD tool for the design and investigation of proposed device structures.